\documentclass[twocolumn,prl,showpacs,preprintnumbers,amsmath,amssymb]{revtex4}

\usepackage{graphicx}
\usepackage[usenames]{color}
\begin{document}

\title{Optically induced rotation of a quantum dot exciton spin}
\author{E.~Poem}
\affiliation{Department of physics, The Technion - Israel institute
of technology, Haifa, 32000, Israel}
\author{O.~Kenneth}
\affiliation{Department of physics, The Technion - Israel institute
of technology, Haifa, 32000, Israel}
\author{Y.~Kodriano}
\affiliation{Department of physics, The Technion - Israel institute
of technology, Haifa, 32000, Israel}
\author{Y.~Benny}
\affiliation{Department of physics, The Technion - Israel institute
of technology, Haifa, 32000, Israel}
\author{S.~Khatsevich}
\affiliation{Department of physics, The Technion - Israel institute
of technology, Haifa, 32000, Israel}

\author{J.~E.~Avron}
\affiliation{Department of physics, The Technion - Israel institute
of technology, Haifa, 32000, Israel}
\author{D.~Gershoni}
\affiliation{Department of physics, The Technion - Israel institute
of technology, Haifa, 32000, Israel}
\email{dg@physics.technion.ac.il}

\date{\today}

\begin{abstract}
We demonstrate control over the spin state of a semiconductor
quantum dot exciton using a polarized picosecond laser pulse
slightly detuned from a biexciton resonance. The control pulse
follows an earlier pulse, which generates an exciton and initializes
its spin state as a coherent superposition of its two non-degenerate
eigenstates. The control pulse preferentially couples one component
of the exciton state to the biexciton state, thereby rotating the
exciton's spin direction. We detect the rotation by measuring the
polarization of the exciton spectral line as a function of the
time-difference between the two pulses. We show experimentally and
theoretically how the angle of rotation depends on the detuning of
the second pulse from the biexciton resonance.

\end{abstract}


\maketitle

Coherent manipulation of the quantum states of a physical system is
a critical step towards novel applications in quantum information
processing (QIP). Semiconductor quantum dots (QDs) exhibit
atomic-like energy spectrum and are compatible with modern micro-
and optoelectronics. They are considered as excellent candidates for
forming the building blocks for these future
technologies~\cite{Loss_DiVincenzo}, and they form the best
interface between quantum light and
matter~\cite{Li03,Zrenner_Nature,Simon10}. Physical realizations of
quantum bits (qubits) and logic gates require controlled two level
systems. The spins of QD-confined charge carriers have been proposed
for this task~\cite{Loss_DiVincenzo,Economou06} and recent
experiments successfully demonstrated their state preparation and
control. Preparation was achieved either by electron-hole pair
photogeneration followed by electrical field induced separation
~\cite{Kroutvar04,Young07,Skolnick_control}, or by optical
pumping~\cite{Atature06,Gerardot08,Press08}. Control of the
initiated state was then demonstrated either by the AC-Stark effect,
induced by an optical pulse~\cite{Press08,Berezovsky08}, or by the
accumulation of a geometric phase~\cite{Wu07,Grelich09,Kim10}. These
impressive achievements require a series of optical pulses and the
presence of a strong fixed magnetic field.

A QD confined electron - hole pair (exciton) is the fundamental
optical excitation in QDs, essential for interfacing light qubits
with matter spin qubits. It was also proposed for QIP
realizations~\cite{Troiani_PRB00,Biolatti00}. The spin state of the
optically active (bright) exciton ~\cite{Bonadeo98,Flis01,Skolnick_gate,Kosaka08},
and that of the dark exciton~\cite{Poem10} have been directly
accessed optically. Moreover, we have recently shown that unlike
carriers' spin, the exciton spin can be initiated at any desired
state by a single optical pulse~\cite{Benny10}. Partial control of
the exciton's spin was recently demonstrated~\cite{Boyer10}, albeit
the time required for a single operation was comparable to the
exciton's radiative lifetime ($\sim$1~ns).

Here we demonstrate, for the first time, control over the spin of
the bright exciton using a single, few picosecond long, laser pulse.
The pulse duration, which is two orders of magnitude shorter than
the exciton lifetime, permits many coherent operations. Moreover,
the same control can be applied to the spin of the dark exciton,
whose lifetime is three orders of magnitude
longer~\cite{warburtonAPL}.

\begin{figure}[tbh]
  \includegraphics[width=0.45\textwidth]{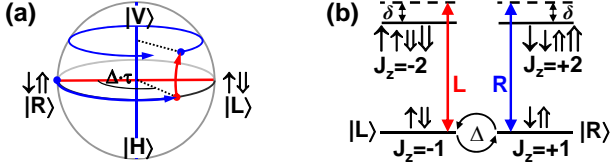}
  \caption{
(a) A Bloch sphere representation of the exciton spin state. The
circle along (above) the equator describes the precession of an
exciton spin initialized at $t=0$ by an R-polarized pulse, for
$t<\tau$ ($t>\tau$), where at $t=\tau$ an R-polarized biexciton
pulse is applied.
(b) Schematic description of the relevant exciton and biexciton
energy levels and the polarization selection rules for the coupling
laser field. The symbol $\uparrow (\Downarrow)$ represents an
electron (heavy hole) with z-direction spin projection $\frac{1}{2}$
($-\frac{3}{2}$). Short (long) symbols represent charge carriers in
their ground (first excited) state.} \label{fig:1}
\end{figure}

Here, we demonstrate this control using a circularly polarized laser
pulse, slightly detuned from a resonance of an excited biexciton
state. The pulse selectively couples one of the exciton spin states
to the biexciton, while leaving the other state unaffected. The
method is conceptually similar to that used recently on ensembles of
charged QDs~\cite{Wu07,Grelich09}. The experiment is conducted as
follows: At first, a right (R) or left (L) circularly polarized
laser pulse tuned into an excitonic resonance of a single
semiconductor quantum dot generates an excited bright exciton. The
spin state of this excited exciton has a total angular momentum
projection of J$_z$=1 ($|R\rangle$) or -1 ($|L\rangle$),
respectively. The excited exciton then non-radiatively relaxes to
the ground-state. As discussed below, the spin is preserved during
this fast relaxation~\cite{Benny10}. This results in the formation
of a coherent superposition of the two non-degenerate ground exciton
eigenstates~\cite{gammon,kulakovskii},
\begin{equation}
|H\rangle=\tfrac{1}{\sqrt{2}}\left(|R\rangle+|L\rangle\right)~;~
|V\rangle=\tfrac{1}{i\sqrt{2}}\left(|R\rangle-|L\rangle\right),
\end{equation}
which by themselves correspond to excitations by horizontally (H) or
vertically (V) linearly polarized pulses, respectively
\cite{Benny10}. Since the eigenstates are non-degenerate, they
evolve at different paces, and the exciton spin precesses in time
between the $|R\rangle$ and the $|L\rangle$
states~\cite{Bonadeo98,Flis01,Skolnick_gate,Benny10} [Fig.~\ref{fig:1}(a)]. Resonant
excitation by an R- polarized pulse results in the following spin
state precession:
\begin{eqnarray}\label{psi_i1}
\nonumber \psi_i(t)=\tfrac{1}{\sqrt{2}}\left(e^{i\Delta\cdot t}|H\rangle+i|V\rangle\right)=a|L\rangle+b|R\rangle~;\\
a=ie^{i\Delta\cdot t/2}\sin(\Delta\cdot t/2);~b=e^{i\Delta\cdot t/2}\cos(\Delta\cdot t/2),
\end{eqnarray}
where $\hbar\Delta$ is the energy difference between the exciton eigenstates.
Then, a second, delayed, circularly polarized pulse is tuned into
(or slightly detuned from) an excited biexciton resonance. As
illustrated in Fig.~\ref{fig:1}(b), this particular resonant level
includes two states, in which the two electron spins are parallel to
each other and they are anti-parallel to the two holes' spins.
The total angular momentum projection of these two
states is J$_z$=$\pm$2.
We note that the biexciton resonance used here, where both the
electrons and the holes are in triplet configurations, is
different than that used in Ref.~\onlinecite{Benny10}, where the
electrons are in a singlet configuration. 

Since an R (L) polarized pulse carries with it angular momentum of 1
(-1), when tuned into the biexciton resonance, it couples only the
J$_z$=1 (-1) exciton state to the J$_z$=2 (-2) biexciton state. The
duration of the laser pulse is much shorter than the period of the
exciton's precession. Hence during the pulse, the coupled exciton -
biexciton states can be safely viewed as an isolated two-level
system \cite{Economou06}.
There is an analytical solution for the dynamics of this system for
the case of hyperbolic-secant pulse
shape~\cite{Economou06,Rosen_Zener,Takagahara10}. The coupling that such a pulse
induces between the relevant exciton and biexciton states is given
by~\cite{Takagahara10},
$C(t)=\mbox{$-\hbar\Omega\cdot$sech$(\sigma t)e^{-i\omega t}$}$,
where $\omega$ is the laser frequency, $\sigma$ is the pulse
bandwidth, and $\Omega$ is the Rabi frequency. If the exciton spin
state just before the second pulse is given by
Eq.~(\ref{psi_i1}), the state after an R-polarized
hyperbolic-secant pulse is given by \cite{Economou06,Takagahara10},
\begin{eqnarray}\label{psi_f}
\nonumber \psi_f=a|L\rangle+bF\left(\alpha,-\alpha,
\gamma,1\right)|R\rangle\\
+b\tfrac{i\alpha}{\gamma}F(\alpha+\gamma,-\alpha+\gamma,1+\gamma,1)|J_z=2\rangle,
\end{eqnarray}
where $F$ is the Gaussian hypergeometric function (also denoted as
$_2F_1$), and
$\alpha=\tfrac{\Omega}{\sigma};\ \
\gamma=\tfrac{1}{2}-\tfrac{i\delta}{2\sigma}$,
where $\delta=\omega-\omega_0$ is the detuning from the resonance
frequency $\omega_0$.
Using known properties of hypergeometric functions
\cite{Takagahara10}, the probability to populate the biexciton if
the time-difference between the two pulses is $\tau$, reads:
\begin{eqnarray}\label{P_XX1}
\nonumber P_{XX}=|\langle J_z=2|\psi_f\rangle|^2=P^0_{XX}|b(\tau)|^2=\\
\mbox{sech}^2\left(\tfrac{\pi\delta}{2\sigma}\right)\sin^2\left(\tfrac{\pi\Omega}{\sigma}\right)\cdot\left[\tfrac{1}{2}+\tfrac{1}{2}\cos(\Delta\cdot\tau)\right].
\end{eqnarray}

A second, non-detuned ($\delta/\sigma=0$) $\pi$-pulse
($\Omega/\sigma=0.5$), coincident in time with the first pulse
($\tau=0$) transfers the entire excitonic population to the
biexciton state~\cite{Economou06}. In general, the
absorption of the second pulse depends on the direction of the
precessing exciton spin relative to the polarization of the light
pulse \cite{Benny10}. Since the intensity of the photoluminescence
(PL) emission from the biexciton spectral lines is a measurement of
the pulse absorption, the emission intensity oscillates as the delay
between the two pulses increases. These oscillations provide an
experimental way to measure the excitonic spin by projecting it onto
directions determined by the polarization of the second
pulse~\cite{Benny10}.
The second pulse affects also the non-transferred excitonic
population. This is because a circularly polarized pulse couples
only one component of the exciton spin state. Thus, the pulse
affects the relative amplitude and phase between the excitonic spin
eigenstates. The change in relative phase can be interpreted as a
``rotation'' of the exciton Bloch sphere about the
$|R\rangle$-$|L\rangle$ axis, and the change in the relative
amplitude as ``squashing'' of the Bloch sphere from the $|L\rangle$
pole towards the $|R\rangle$ pole. The exciton state after such an
operation is generally expressed, up to normalization, by the first two terms on the right
hand side of Eq.~(\ref{psi_f}). In order to detect it, one needs to
measure the spin direction of the exciton after the second pulse. We
show below that the excitonic spin projection on the
$|H\rangle$-$|V\rangle$ axis of the spin Bloch sphere is readily
available by measuring the net polarization of the PL from the
exciton lines.

The idea is schematically described in Fig.1(a): As long as the
exciton's spin is oriented along the Bloch sphere's equator, its
projection on the $|H\rangle$-$|V\rangle$ axis is zero, and both
eigenstates of the exciton are equally populated. In contrast, if
the spin is forced, by the second pulse, to move in a trajectory
which leaves the equator, then the populations of the two
eigenstates are no longer equal. Once the second pulse is turned
off, the exciton spin again precesses around the
$|H\rangle$-$|V\rangle$ eigenstates axis, and the eigenstates
population difference created during the pulse is kept constant.
Since the PL emission is proportional to the probability of
population, variations in the population result in measurable
changes in the PL from the exciton's spectral lines. Clearly, the
normalized difference between the emission intensities of the two
cross-linearly polarized exciton lines is a direct experimental
measurement of the spin projection on the $|H\rangle$-$|V\rangle$
axis of the Bloch sphere. The energy difference between the cross
linearly polarized components is larger than their spectral widths.
Therefore, their intensities can be simultaneously measured using a
circular polarizer in front of the monochromator.
%
%

The excitonic PL emission in our experiment is not temporally
resolved. Therefore, it also contains contribution from the
biexciton population [Eq.~(\ref{P_XX1})], which decays
\emph{incoherently} into excitonic population. However, due to the
polarization selection rules [Fig.~\ref{fig:1}(b)], these incoherent
excitons equally populate the $|H\rangle$ or $|V\rangle$
eigenstates, and therefore do not contribute to the eigenstates
population difference. The difference can thus be calculated
directly from the exciton state immediately after the second pulse,
Eq.~(\ref{psi_f}). With Eq.~(\ref{psi_i1})
and using properties of hypergeometric functions \cite{Takagahara10}
one obtains,
\begin{eqnarray}\label{X_diff1}
D_{VH}=|\langle V|\psi_f\rangle|^2-|\langle H|\psi_f\rangle|^2=&\\
\nonumber -2\mbox{Re}\left[a(\tau)b(\tau)^*F\left(\alpha,-\alpha,\gamma,1\right)^*\right]=&D^0_{VH}\sin\left(\Delta\cdot\tau\right);\\
\label{X_diff2}
D^0_{VH}=\mbox{Im\small{$\left[\frac{\Gamma^2\left(\frac{1}{2}+\frac{i\delta}{2\sigma}
\right)}{\Gamma\left(\frac{1}{2}+\frac{i\delta}{2\sigma}+\frac{\Omega}{\sigma}
\right)\Gamma\left(\frac{1}{2}+\frac{i\delta}{2\sigma}-\frac{\Omega}{\sigma}
\right)}\right]$}},
\end{eqnarray}
$\Gamma(z)$ being the Gamma function. Eq.~(\ref{X_diff1}) shows that
the oscillations in $D_{VH}$ have the same frequency as those of the
biexcitonic population, Eq.~(\ref{P_XX1}), but they are shifted in
phase by $\pi$/2. The amplitude of the oscillations,
Eq.~(\ref{X_diff2}), depends on the pulse intensity ($\Omega$), its
bandwidth ($\sigma$) and its detuning ($\delta$). In particular, the
sign of the amplitude is given by the sign of the detuning, and on
resonance the amplitude vanishes. By using
Eqs.~(\ref{psi_f})-(\ref{X_diff2}), the angle of the induced
rotation is given by
\begin{equation}\label{angle}
\theta=\sin^{-1}\left(\mbox{\small{$D^0_{VH}/\sqrt{1-P^0_{XX}}$}}\right).
\end{equation}
A larger-than-$\pi$ pulse
permits any rotation angle.

The studied sample contains one layer of strain-induced InGaAs QDs
embedded in the center of a one optical wavelength (in matter)
microcavity that enhances the PL collection efficiency. For the
optical measurements the sample was placed inside a tube immersed in
liquid Helium, maintaining sample temperature of 4.2K. A spatially
isolated single QD from a low-density area of the sample was excited
with two dye lasers, synchronously pumped by a frequency doubled
Nd:YVO$_4$ passively mode-locked pulsed laser. The dye laser's pulse
temporal full-width-at-half-maximum (FWHM) was measured to be 9~ps.
An in-situ microscope objective was used both to focus the exciting
beams onto the sample, and to collect the light emitted from it. The
collected light was projected upon a desired polarization, dispersed
by a 1~meter monochromator, and then detected by either an
electrically-cooled charge-coupled-device array detector or by
a single channel, single photon, silicon avalanche photodetector.
The system provides spectral resolution of about 10~$\mu$eV.
One dye laser was tuned to an excited exciton resonance, 29~meV
above the ground-state exciton emission line. Excitation in this
resonance results in a high ($>90\%$) degree of linear polarization
memory of the exciton emission line. Moreover, the width of this
resonance, which is greater than 500~$\mu$eV, is much larger than
its polarization splitting of 60~$\mu$eV \cite{Benny10}. Therefore,
relaxation to the ground exciton state occurs before any appreciable
dephasing or rotation of the spin state can occur, and the spin
state of the excited exciton, determined by the polarization of the
exciting laser, is preserved during this fast
relaxation~\cite{Poem10,Benny10}. The other dye laser was tuned into
(or slightly detuned from) an excited biexciton resonance, at
33.7~meV above the exciton doublet. Using detailed polarization
sensitive spectroscopic studies~\cite{Benny11} and a
many-carrier theoretical model~\cite{Poem07} we unambiguously
identified this resonance as the \mbox{J$_z$=$\pm$2} biexciton
states described in Fig.~\ref{fig:1}(b).
\begin{figure}[tbh]
  \includegraphics[width=0.48\textwidth]{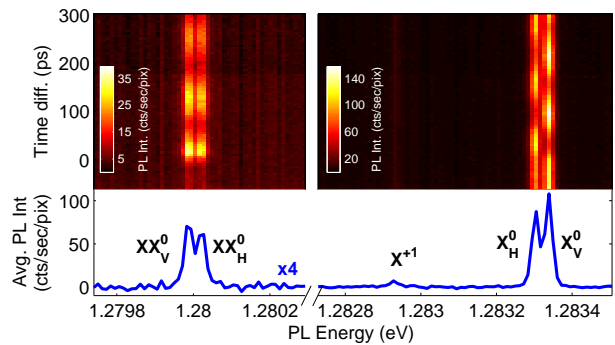}
  \caption{R-polarized PL intensity as a function of the photon energy and the time difference
  between two R-polarized laser pulses. The first is tuned into the exciton resonance and the second is
detuned by \mbox{-63~$\mu$eV} from the biexciton resonance. The
temporally averaged spectrum is shown at the bottom.}
  \label{fig:2}
\end{figure}

In Fig.~\ref{fig:2} we present the R polarized PL intensity from the
QD vs. the PL energy and the time difference ($\tau$) between the
first pulse, tuned into the exciton resonance, and the second one,
detuned by -63~$\mu$eV from the biexciton resonance. Both lasers
were R polarized. The higher (lower) energy doublet is the emission
from the ground state exciton (biexciton). The oscillations of the
PL from the biexciton reflect the precession of the exciton spin
state, as initialized by the first pulse~\cite{Benny10}. This
behavior is described by Eq.~(\ref{P_XX1}). The oscillations of the
PL from the two excitonic components, reflect the variations of the
exciton spin projection on the eigenstates axis, induced by the
detuned second pulse, as described by Eq.~(\ref{X_diff1}). The line
at 1.28293 eV is due to the positively charged exciton. Its PL does
not oscillate with $\tau$.
\begin{figure}[tbh]
  \includegraphics[width=0.47\textwidth]{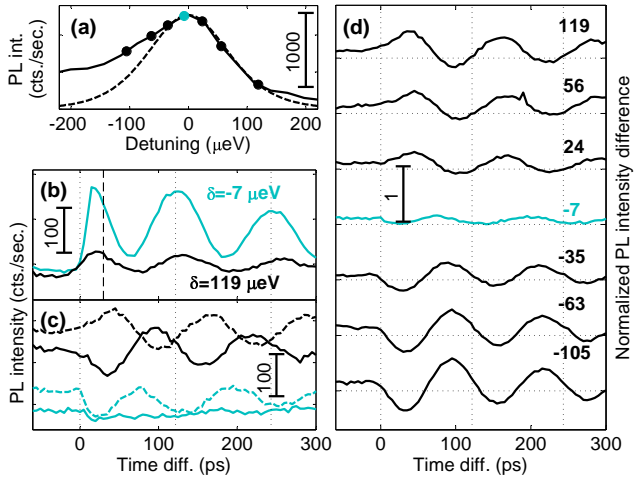}
  \caption{(a) Solid line - measured biexciton PL intensity vs.
  the second pulse detuning, at
  $\tau=30$~ps. The circles represent the specific
  energies used for the measurements presented in (b)-(d). The
  dash line is the calculated $P_{XX}$ [Eq.~(\ref{P_XX1})], for a pulse
of 9~ps FWHM ($\hbar\sigma=145~\mu$eV).
  (b) Biexciton intensity vs. $\tau$ close to (light blue) and
  far from (black) resonance. The vertical dash line denotes
  $\tau=30$~ps. (c) Solid (dash) line denotes PL intensity
  of the $H$ ($V$) component of the exciton doublet vs. $\tau$, close to and far from
  resonance, as in (b). The curves are vertically shifted for clarity.
  (d) The differences between the PL from the $V$ and $H$ components, normalized by their
  sum at negative $\tau$, for the energies marked by circles in (a) and specified
  in $\mu$eV next to each curve.
  Vertical dotted lines present integer spin precession periods \mbox{T=$h$/(34
  $\mu$eV)=122~ps}.
  }
  \label{fig:3}
\end{figure}

\begin{figure}[tbh]
  \includegraphics[width=0.48\textwidth]{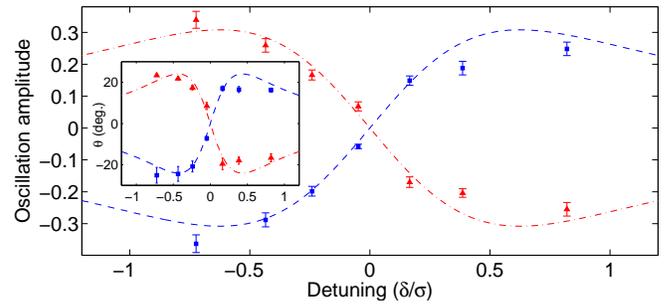}
  \caption{Measured (symbols) and calculated (lines)
oscillation amplitudes of the exciton polarization vs. the detuning
$\delta/\sigma$, where $\hbar\sigma=145\mu$eV and
$\Omega/\sigma=0.35$ (0.7$\pi$-pulse). Blue (red) color describes
co- (cross-) circularly polarized pulses.
  Inset: The rotation angles, $\theta$, vs. the detuning for a 0.7$\pi$ pulse.}
  \label{fig:4}
\end{figure}

Fig.~\ref{fig:3}(a) presents a PL-excitation spectrum of the
biexciton line, for 30~ps difference between the first and second
pulse. The intensity of the second pulse was tuned slightly below
population inversion at resonance excitation. At this particular
intensity the emission intensity from both the exciton and biexciton
lines is optimized. This allows simultaneous testing of
Eqs.~(\ref{P_XX1}) and (\ref{X_diff1}). The dashed line represents
the calculated biexciton population [Eq.~(\ref{P_XX1})], for a 9~ps
FWHM hyperbolic-secant pulse ($\hbar\sigma$=145~$\mu$eV). The
deviation of the measured intensity from the theoretical line at low
energies is due to a near-by \mbox{J$_z$=0} excited biexciton
resonance~\cite{Benny11}.
We note that the width of the J$_z$=$\pm$2 biexciton resonance is
completely determined by the spectral width of the laser. This
indicates that any dephasing and relaxation processes are
significantly slower than the pulse duration. Indeed, in cw
excitation the spectral width of this resonance is significantly
narrower. This implies that the J$_z$=$\pm$2 biexciton remains
coherent during the second pulse.
In Fig.~\ref{fig:3}(b) we present the intensity of the PL from the
biexciton lines as a function of $\tau$. Black (blue) line presents
off (almost on) biexcitonic resonant excitation. In both cases, the
evolution is cosinusoidal, as in Eq.~(\ref{P_XX1}). In
Fig.~\ref{fig:3}(c) we present the intensity of the PL from the
exciton lines as a function of $\tau$, for both off and almost on
resonant excitation as in (b).
Solid (dash) line denotes the PL from the H- (V-) polarized
component of the excitonic doublet. In Fig.~\ref{fig:3}(d) we
present the difference between the PL intensities from these two
cross-linearly polarized components, normalized by their sum at
negative delay time (before the second pulse). The oscillations are
sinusoidal in time, as in Eq.~(\ref{X_diff1}). Their amplitude
depends on the detuning from resonance. below (above) resonance, the
amplitude is negative (positive) and on resonance it vanishes, as
expected from Eq.~(\ref{X_diff2}).
This dependence is summarized in Fig.~\ref{fig:4}, which presents
the amplitudes of the measured oscillations in the exciton
polarization, vs. the normalized detuning, $\delta/\sigma$. The
lines present the amplitude calculated by Eq.~(\ref{X_diff2}), for
$\Omega/\sigma$=0.35 (0.7-$\pi$ pulse), slightly below inversion.
The figure describes co- (blue) and cross- (red) circularly
polarized laser pulses. The measured and calculated spin rotation
angles [Eq.~(\ref{angle})] vs. the detuning are presented in the
inset to Fig.~\ref{fig:4}. The agreement between the measured
rotations and the theoretically calculated ones (which assume no
dephasing) indicates a close to unity rotation fidelity.

In summary, we demonstrated that the polarization of the QD exciton
spin can be rotated by a single, short optical pulse. We showed that
the rotation can be directly detected by monitoring the intensities
of the two components of the excitonic spectral line.

The support of the US-Israel binational science foundation (BSF),
the Israeli science foundation (ISF), the ministry of science and
technology (MOST) and that of the Technion's RBNI are gratefully
acknowledged.


\end{document}